# From user requirements to UML class diagram

*Hatem Herchi, **Wahiba Ben Abdessalem
*Department of computer science, High Institute of Management of Tunis, Tunisia
**Department of computer science, High Institute of Management of Tunis, Tunisia

*Abstract*—The transition from user requirements to UML diagrams is a difficult task for the designer especially when he handles large texts expressing these needs. Modeling class Diagram must be performed frequently, even during the development of a simple application. This paper proposes an approach to facilitate class diagram extraction from textual requirements using NLP techniques and domain ontology.

## I. INTRODUCTION

UML class diagrams are the main core of OO analysis and design where other models are derived.

Sometimes, due to the time and cost factors involved in reconstruction of the class diagram, designers modify them directly. As a result, inconsistency can be introduced in this case between user requirements documents and the formal models, which in turn causes not only serious problems in the application maintenance phase, but also affects the prototyping and reusability of new software with similar requirements.

To analyze a given text, the most Natural Language Processing (NLP) systems are based on the following levels: Morphological level, lexical level, syntactic level, semantic level, discourse level and pragmatic level [2]. Domain ontology has also been widely used to improve the efficiency of concepts identification. It models a specific domain, which stands for a part of the world. Using this kind of ontologies, organizations, enterprises or communities describe the concepts in their domain, the relations between those concepts, and obviously the instances that are the actual things that fill its structure.

The main goal of the present work is to investigate how Natural Language Processing techniques and Domain Ontology can be exploited to support the Object-Oriented Analysis process. This study must accept, as an input, textual data expressed in natural language and representing the user needs then identify the classes' names, their attributes and associations between them in order to classify them in a structured XML file.

The rest of this paper is organized as follows: section 2 analyzes related works on natural language based object oriented analysis and modeling. Section 3 gives an overview of GATE API. Section 4 described in detail our implemented system. Section 5 discusses the evaluation methodology and experimental results and the final section presents conclusions and futures works.

## II. RELATED WORKS

First, Ambriola and Gervasi [3] propose a framework to automatically transform user's requirements into different models such as Data Flow graphs (DFG), entity-relationship diagrams, or even UML diagrams. Indeed, the system takes as an input the problem's description written in natural language, and then a domain based parser, called CICO, and is used to extract some facts from them. These facts are then processed by the other tools for analysis or graphical representation.

Zhou and Zhou [4] present and implement a system that automates class diagram generation from free-text requirement documents. The approach firstly applies NLP techniques to understand written requirements and then uses domain ontology to improve the performance of class identification. In fact, this methodology extracts candidate classes using part of speech tagger, a link grammar parser, parallel structure and linguistic patterns. The output is then refined using domain ontology.

Mich L. [5] proposed a natural language processing tool named LOLITA (Large-scale Object-based Language Interactor, translator and Analyser). It is used to pre-process the user requirements; it includes all the tasks for natural language anal
ysis. It's built around SemNet which is a semantic graph that contains a large number of object and event nodes used to bridge the gap between object diagrams and requirements. Indeed, this approach considers nouns as objects and use links to find relationships between objects. LOLITA extract only objects from natural language and it cannot distinguish between classes, attributes and objects.

## III. GATE OVERVIEW

GATE was implemented by the University of Sheffield in 1995 and it was released for the first time in 1996. This open source framework is developed using the Java programming language. Therefore, it can run on any platform including the Java Virtual Machine. It is used for developing software components that process natural language.[1]

We resort to **GATE** because it has proved its efficiency. It can provides a set of natural language analysis tools which can take English language text input and give as a result the base forms of words, their parts of speech, etc., and mark up the structure of sentences in terms of phrases and word dependencies, and mention which noun phrases refer to the same entities. In other words, GATE offers the foundational building blocks for higher level text understanding applications.

---

[1] http://gate.ac.uk/

GATE has an information extraction (IE) system called

IV. DC-BUILDER SYSTEM DESCRIPTION

The following figure (fig. 1) describes DC-Builder architecture.

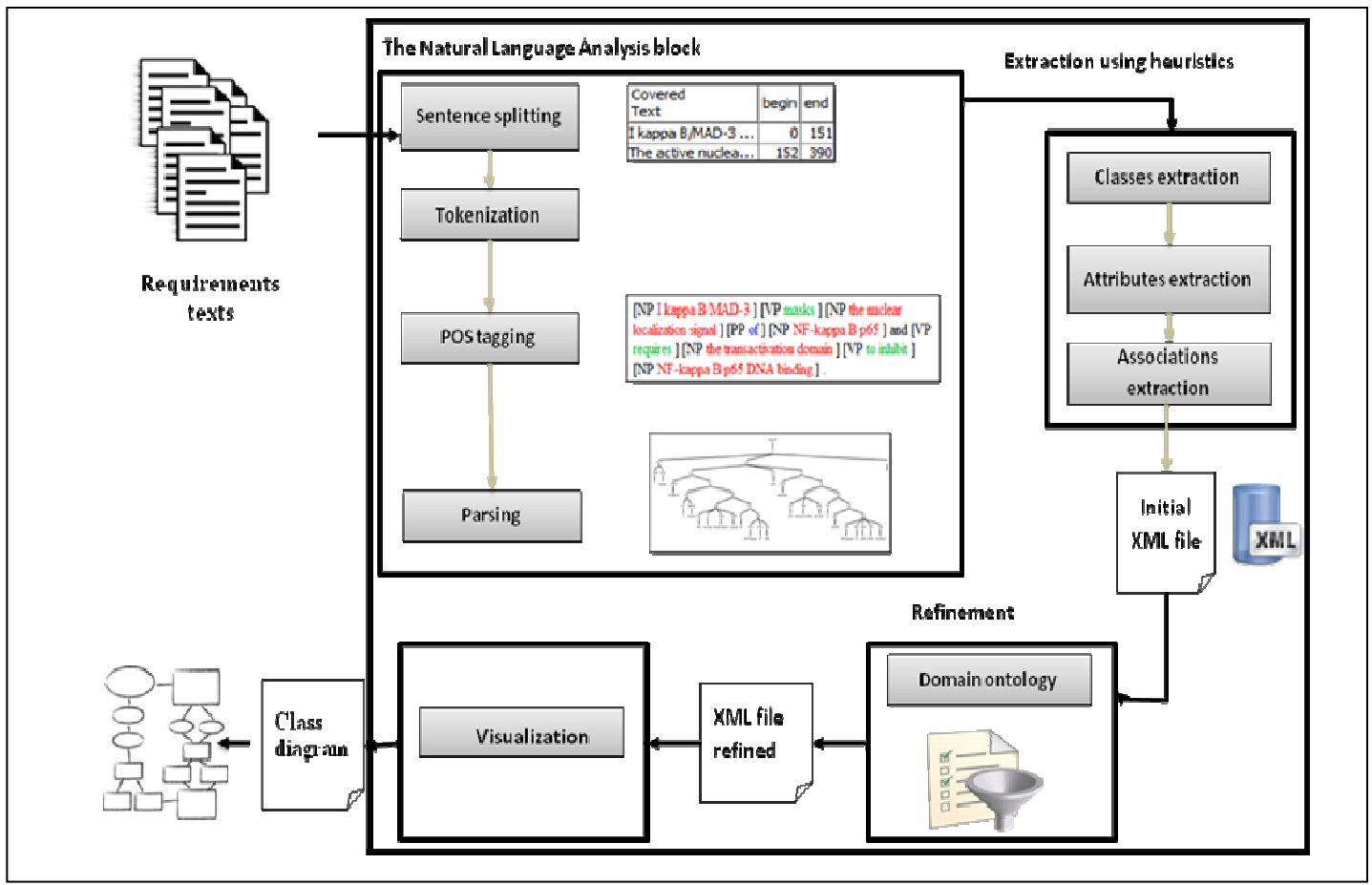

ANNIE (A Nearly-New Information Extraction System) which contains several language processing:

- **Sentence splitter:** the sentence splitter separates each sentence from the input string and returns a list of strings.
- **Tokenizer:** the tokenizer takes each sentence as an input and splits them into tokens such as numbers, words and punctuation.
- **Parts of speech (POS tagger):** It is used to perform the process of marking up the words in a text as corresponding to a particular part of speech.
- **Syntactic parser:** sequences of words are transformed into structures that indicate how the sentence's units relate to each other. This step helps us in identifying the main parts in a given sentence such as object, subject, verb…etc.

Figure 1. DC-Builder System architecture

*A. Natural Language Analysis block*

The natural language analysis block processes the requirements descriptions submitted by the user using the framework GATE, and specially: Sentence splitter

*B. UML concepts extraction using heuristics*

This section focuses in heuristics and their application to improve the generation of OO concepts from natural language texts: we strongly stress that we will not decorticate all the existing heuristics which were published before, but we will focus only in the relevant ones that can serve us.

Usually, candidate classes can be extracted by considering the noun phrases in the requirements text. Candidate relationships can be found in the same way by considering verb phrases. For example, by analyzing the sentence "a doctor gives medicines to the patient" we can find out three candidate classes (doctor, medicines, and patient) and one candidate relationship (gives).

In this context, heuristics can play a fundamental role to facilitate such task. In fact, heuristics, guided by common sense, provides good but not necessarily optimal solutions to many difficult problems where precise and pertinent algorithmic solutions are not available such as those treated in this thesis.

Given a parts-of-speech and their functions in sentences, Chen [6] proposed eleven rules in order to translate NL requirements description written in English from natural language (English) to an entity-relationship diagram. The proposal of Chen seems to be the first attempt using linguistic concepts to conceptual modeling. These various rules reflect Chen's experience and knowledge of the ER meta-model.

The rules described below will be used to facilitate the extraction of classes' names for the future diagram :

*Rule 1: "All nouns are converted to entity types" [7]*

We can conclude that all nouns can be mapped to classes' names; we mean by nouns all type of nouns such as common nouns, collective nouns, proper nouns, mass nouns and count nouns.

*Rule 2: "A gerund may indicate an entity type which is converted from a relationship type" [6]*

Firstly, A gerund can be defined as a noun which consists of a verb and an "ing". It is often called an -ing word or a verbal noun.

*Rule 3: a specialization's relationship between entities: sentence's structure "is a" can relate two nouns A and B to one another. [8]*

*Rule 4: A noun such as "database", "record", "system", "company", "information", "organization" and "detail" may not be considered as a relevant candidate for an entity type since it shows the business environment and logically have to be not included in the entity's category [9].*

*Rule 5: every proper noun (Person name, Location name …) is ignored to be a class.*

This rule can help us to perform a partial filtering in order to obtain an accurate set of classes' names for the future class diagram.

For attributes extraction, we enumerate some heuristics:

*Rule 6: A noun such as "vehicle_number", "group_no", "person_id" and "room_type" may refer to an attribute type [10*

*Rule 7: The genitive case, also called possessive case, often shows ownership. Hence, it can be used to extract attributes [7]*

*Rule 8: If consecutive nouns are present, check the last noun. If it is not one of the words in set S where S = {number, no, code, date, type, volume, birth, id, address, name}, most likely it is an entity type. Else it may indicate an attribute type. [9]*

*Rule 9: A noun phrase succeeding the "has/have" verb phrase may indicate the presence of attribute types [10]*

For associations' extraction, we use Three heuristics:

*Rule 10: A transitive verb can be a candidate for relationship type [6]*

Transitive verb, in syntax, is a one that requires an object to complete its meaning. This verb may be considered as a candidate for an association.

*Rule 11: A verb followed by a preposition such as "in", "on", 'to" and "by" can indicate a relationship type[9]".*

*Rule 12: "if a verb is equal to one of the following list {"include", "involve", "consists of", contain, "comprise", "divided to", "embrace"}, therefore, this relationship can be aggregation or composition".*

In this section has presented background information about the various linguistic rules that can support the UML concepts extraction. As an input, the concerned module produces an initial XML file that should be refined.

C. *Refinement using domain ontology*

The previous module produces an initial model, in XML format, including concepts related to classes, attributes and associations. This model can contain erroneous elements which should be treated. Indeed, the built ontology will help us to eliminate irrelevant elements, and then keep only those which will be used to construct the final class diagram. The unimportant elements are simply detected and those which don't pertain to the ontology will automatically deleted

D. *DC-Builder implementation*

We developed the DC-Builder System using Eclipse IDE (Version 3.7.1). Indeed, Eclipse has a special feature by the fact that its architecture is developed around plug-ins. This concept can provide a mechanism to extend the platform functionalities and allow users to integrate components according to their needs. DC-Builder can open textual requirements from various sources such as text files (TXT), words documents (DOC) and rich text files (RTF).

V. EVALUATION

We test case studies published in Information Systems and Object-Oriented Analysis books.

The results of these case studies were used to calculate recall, precision and overgeneration. We compare our tool with CM-Builder,values as shown in table 1:

TABLE I. EVALUATION RESULTS

| | CM-Builder | DC-Builder |
|---|---|---|
| Recall | 80.5 % | 83% |
| Precision | 88 % | 93% |
| Over-generation | 51.5 % | 57 % |

In addition, the various tools' functionalities (if available, is automated or user involved) are also compared with DC-Builder as shown in Table 2:

TABLE II. EVALUATION TOOLS' FUNCTIONALITIES

| Support | CM-Builder | LIDA | GOOAL | NL-OOML | DC-Builder |
|---|---|---|---|---|---|
| Classes | Yes | User | Yes | Yes | Yes |
| attributes | Yes | User | Yes | Yes | Yes |
| Methods | No | User | Yes | Yes | NO |
| Associations | Yes | User | Semi-NL | No | Yes |
| Multiplicity | Yes | User | No | No | No |
| Aggregation | No | No | No | No | Yes |
| Generalization | No | No | No | No | Yes |
| instances | No | No | No | No | No |

The evaluation shows that besides DC-Builder, there are few tools those can extract information such as aggregations, generalizations from NL requirement. Thus, the results of this performance evaluation are very motivating and support both the potential of this technology and the approach adopted in this paper.

## VI. CONCLUSIONS

This paper aims to automate the analysis of software requirements documents using Natural Language Processing techniques in order to build initial UML Class Models. In this context, a NLP-based tool, called Diagram Class Builder (DC-Builder), is developed to achieve this objective.

We resort to GATE API to process the input scenario. Then we use some linguistic rules (heuristics) to find out UML concepts such as classes names, their attributes, associations. As a result, we obtain an initial XML file that should be refined.

To achieve this goal, our approach integrates domain ontology which contains core domain knowledge. In our context, the domain ontology seems to be the right way to improve the quality of outputs and the performance of concepts identification. Indeed, it feeds the system relevant classes and their attributes, nevertheless, classes and attributes are not limited to those in the ontology. The results produced by the Diagram Class Builder for two different case studies (A Library Information System and An Automatic Teller Machine) were analyzed in detail, and the system's evaluation is based on comparing the outputs with class models which are manually created. Obtained results were motivating and have demonstrated the benefits of our approach. However there are several extensions that can be added to improve the system's performance:

- The heuristic rules cited in this thesis are not exhaustive. There are certain sentence structures that are not covered by the proposed heuristics.

- We should improve text analysis algorithms in order to generate complex UML diagrams such as activity and sequence diagrams. Indeed, our approach focuses on the extraction of Class diagram elements (classes, attributes, and associations) that shows the static aspect of the analyzed system. The same process can be done to model the dynamic aspect of the system.

In the future works, we will cooperate to solve the described challenges and improve the NL-based class diagram building technology.